\documentclass[10pt,printer]{aa}
\usepackage{graphicx}
\usepackage{natbib}
\bibpunct{(}{)}{;}{a}{}{,}
\begin{document}
\title{Physical and dynamical characterisation of low $\Delta$V NEA (190491)
{2000 $\mbox{FJ}_{10}$}\thanks{based on observations made with the
Southern African Large Telescope (SALT)}}
\author {A.~A.~Christou\inst{1} \and
T.~Kwiatkowski \inst{2} \and M.~Butkiewicz \inst{2} 
\and A.~Gulbis \inst{3} \and C.~W.~Hergenrother \inst{4}
\and S.~Duddy \inst{5} \and A.~Fitzsimmons \inst{6}}

\institute{
           Armagh Observatory, College Hill,
           Armagh BT61 9DG, UK
           e-mail: aac@arm.ac.uk
      \and Astronomical Observatory Institute,
           Faculty of Physics, A.~Mickiewicz University,
           S{\l}oneczna 36, 60-286 Pozna\'{n}, Poland 
   \and Southern African Large Telescope and 
            South African Astronomical Observatory,
           Observatory Road,
           Observatory,
           7925, South Africa 
      \and Lunar and Planetary Laboratory,
              University of Arizona, 
              Tucson AZ 85721-0092, USA
      \and Centre for Astrophysics and Planetary Science, 
               School of Physical Sciences, 
              University of Kent, Canterbury CT2 7NH, UK
      \and Astrophysics Research Centre, School of Mathematics and Physics, 
           Queen's University Belfast, Belfast BT7 1NN, UK
          }
\date{Received 02 Aug 2012 / Accepted Oct 2012}
\abstract
{}
{We investigated the physical properties and dynamical evolution of Near Earth Asteroid (NEA) (190491) 2000 $\mbox{FJ}_{10}$ in order to assess 
the suitability of this accessible NEA as a space mission target.}
{Photometry and colour determination were carried out with the 1.54 m Kuiper Telescope (Mt Bigelow, USA) and the 10 m Southern African Large Telescope 
(SALT; Sutherland, South Africa) during the object's recent favourable apparition in 2011-12. 
During the earlier 2008 apparition, a spectrum of the object in the 6000-9000 Angstrom region was obtained with the 
4.2 m William Herschel Telescope (WHT; Canary Ils, Spain). Interpretation of the observational results was aided by numerical 
simulations of 1000 dynamical clones of 2000 $\mbox{FJ}_{10}$ up to $10^{6}$ yr in the past and in the future.}
{The asteroid's spectrum and colours determined by our observations suggest a taxonomic classification within the S-complex although 
other classifications (V, D, E, M, P) cannot be ruled out. 
On this evidence, it is unlikely to be a primitive, relatively unaltered remnant from the early history of the solar system and thus a low priority target 
for robotic sample return. Our photometry placed a lower bound of 2 hrs to the asteroid's rotation period.  
Its absolute magnitude was estimated to be $21.54 \pm 0.1$ which, for a typical S-complex albedo, translates into a diameter 
of $130\pm20$ m. Our dynamical simulations show that it has likely been an Amor for the past $10^{5}$ yr. Although currently not Earth-crossing, it will likely become so during the period $50 - 100$ kyr in the future. It may have arrived from the inner or central Main Belt $> 1$ Myr ago as a former member of a low-inclination S-class asteroid family. Its relatively slow rotation and large size make it a suitable destination
for a human mission. We show that ballistic Earth-190491-Earth transfer trajectories with $\Delta$V $< 2$  km $\mathrm{s}^{-1}$ at the asteroid exist between 2052 and 2061.}
{}
\titlerunning{Physical and dynamical characterisation of  2000 $\mbox{FJ}_{10}$}
\authorrunning{Christou, et al.}
\maketitle
\section{Introduction}

 The population of near-Earth asteroids (NEAs) contains a small
fraction of objects in low inclination, low-eccentricity orbits similar to
the Earth's.  These NEAs are considered attractive targets for in situ
investigation by robots or humans \citep{Hasegawa.et.al2008,Abell.et.al2009,Michel.et.al2009,Lauretta.et.al2010, Elvis.et.al2011}. 
However, the attractiveness of individual objects as targets for either
robotic or human missions is mired by the currently poor knowledge of their
orbits and physical properties.  Many have been observed only on a single
apparition resulting in large projected uncertainties in their future
position.  In addition, knowledge of properties that are important from an
operational as well as a scientific standpoint - size, shape, surface
roughness, rotational state and spectral type - ranges from poor to
non-existent.  Part of the problem stems from their Earth-like orbits; slow
keplerian shear generally places them beyond the reach of Earth-based
observatories except during the few months that they spend in proximity to
the Earth every decade or so.  The situation is also not helped by their
small sizes, typically a few tens of metres, so even when near the Earth
their study is the exclusive purvue of large-aperture instruments. 
Primitive NEA taxonomies - those belonging to classes B, C, D, and P - are
preferred as mission targets for robotic sample return as they are thought to be relatively unaltered
relics of the early solar system \cite[eg][]{Michel.et.al2009}.  In addition, fast
rotators are unsuitable as mission targets due to the added complexity of
operations in close proximity to such objects.

This paper reports on a study of NEA (190491) 2000 $\mbox{FJ}_{10}$, an object that is
accessible from the Earth (Section \ref{sec:WhyFJ10}).  A programme of observations of 190491 was
carried out with the 4.2 m William Herschel Telescope (WHT), the 10 m Southern African
Large Telescope (SALT) and the 1.5~m Kuiper telescope (Section \ref{sec:Obs}) aiming to constrain
its taxonomic type, size, and rotational state (Section \ref{sec:Phys}).  Those were combined with numerical simulations of 190491's 
orbital evolution (Section \ref{sec:Orbit}) to provide context for the observational characterisation and 
help us draw conclusions on the object's likely origin (Section \ref{sec:Origin}). Its accessibility from the Earth was quantified
by constructing direct, two-way keplerian trajectories between the Earth and the asteroid (Section \ref{sec:Access}).
A summary of our findings is provided in Section \ref{sec:Summary}.

\section{The Asteroid}
\label{sec:WhyFJ10}

(190491) 2000 $\mbox{FJ}_{10}$ was discovered by the Spacewatch survey on 25 March 2000. 
Based on its orbital parameters ($a=1.32$ AU, $e=0.23$, $i = 5^{\circ}$) it is
classified as an Amor Near Earth Asteroid (NEA).  Its absolute magnitude
$H=20.9$ implies a diameter ranging from 110-390 m for an albedo range
0.05-0.5.  Its Earth Minimum Orbit Intersection Distance (MOID) is 0.055 AU,
slightly higher than the threshold of 0.05 AU for classifying it as a
Potentially Hazardous Asteroid (PHA).  It is one of the most accessible
spacecraft targets, ranking 124th out of 8857 object entries in the list of
Near Earth Asteroid $\Delta$V for spacecraft rendezvous\footnote{
http://echo.jpl.nasa.gov/$\sim$lance/delta\_v/
delta\_v.rendezvous.html} as
of May 2012. Its $\Delta$V as stated in that list is 
$4.567$ $\mathrm{km}$ $\mathrm{s}^{-1}$,
slightly above the threshold that would classify it an Ultra Low Delta V (ULDV)
object \citep{Elvis.et.al2011}.  However, its absolute magnitude is the
second-brightest within those objects that precede it, 
the first being the ULDV NEO (89136) 2001 US16 with $H=20.2$ and
$\Delta$V $= 4.428$ km $\mathrm{s}^{-1}$.

\section{Observations and Data reduction}
\label{sec:Obs}
\subsection{SALT}

Located at SAAO in South Africa, the 10 m SALT is based on the Hobby-Eberly Telescope (Texas, USA) with a
payload moving during the observations above the stationary 10~m spherical
mirror \cite[][ and references therein]{Kwiatkowski.et.al2010}.
It can access objects in the declination range from $\delta =-75{\degr}$ 
to $\delta=+10{\degr}$ when they enter the annular region on the
sky located between zenith distances from $48{\degr}$ to $59{\degr}$. 
SALT works solely in the queue-scheduling 
mode, in which the exact time of the observations is not known in advance.

Because of its construction, during observations the telescope's pupil
continuously changes making it impossible to perform all-sky photometry.
For the same reason, twilight flat fields cannot be used in the
photometric reduction. Instead, night sky flat fields derived directly from
the science frames have to be used.

We observed 2000 $\mbox{FJ}_{10}$ with SALT during September 2011, on the first
month of the facility's normal operation after an extended period of commissioning. 
The instrument of choice was the SALT imaging camera \citep[SALTICAM;][]{ODonogue.et.al.2003}, a mosaic of two CCDs, each with two readout amplifiers. 

The aspect data and the observing log are
provided in Table~\ref{tab:AspectData}. During that period the asteroid was
located within a star field covered by the Sloan Digital Sky Survey
(hereafter SDSS), which allowed us
to calibrate our photometry with SDSS standard stars. 
Under normal operations (the case for the observations reported here), there is an effort to keep target and comparison sources on the same chip so as to ease data analysis. 
This limited our Field of View (hereafter FoV) but still left us with enough SDSS stars for comparison. 

In order to determine colour indices we used the Sloan g, r and i filters. 
Because of the significant fringing in the infrared, the z filter was not used. 
As the lower bound of the range of the possible effective diameters of 2000
$\mbox{FJ}_{10}$ was 110~m, its rotation period could be as short as 10~min \citep[eg][]{HergenrotherWhiteley2011}. To check
that, the first run on 15 Sep, which lasted 20 min, was done with the r filter
only.  A preliminary reduction showed all photometric data points to be within the $\pm
0.05\,\,\mathrm{mag}$ range.  The next run on 22 Sep lasted 50 min.  It was
executed with all three filters, in the following sequence: 10 $\times$~r, 3
$\times$~g, r, 3 $\times$~g, r, 3 $\times$~g, r, 3 $\times$~i, r, 3
$\times$~i.  Sequencing the observations in this way allowed us to use r exposures to search for
brightness variations, and -- if necessary -- reduce the magnitudes
obtained in the other two filters to the same reference level. 

Data reduction was done in two stages. First, the CCD frames were
corrected for cross-talk and bias using the
specialised PySALT package \citep{Crawford.et.al.2010}.  Next, following 
\citet{KniazevVaisanen2011}, a flat-field correction was
carried out using night sky flat-fields obtained from the science frames themselves.
 
An illumination pattern for each frame was created by
removing the stars with a median filter and fitting a polynomial to the
remaining background sky. The original frames were then divided by these
flat-fields.  During the observations the telescope was dithered every 3 exposures 
so that the stars did not occupy the same positions within the FoV. 
This allowed us to use all frames in a given filter -  already
corrected for the low frequency pattern - to produce a global, second-stage
flat-field by median combining the images.  The obtained
flat-field mapped the constant pixel-to-pixel variations within the FoV.  
Finally, all images in a given filter were divided by the global
flat-field frame.  

The instrumental magnitudes of the SDSS stars and of the
asteroid were then measured with standard aperture photometry.  Both the
flat-field corrections and the photometry were done with the STARLINK
package, maintained at the Joint Astronomy Centre of Mauna Kea
Observatory.

A problem in calibrating the instrumental magnitudes to the SDSS
standard scale was the unknown transformation between the photometric
systems of SALT and SDSS.  While this would require a detailed analysis, we
instead performed a simple, first-order check with the available data, obtained
through the SALT g$_{S}$, r$_{S}$, and i$_{S}$ filters .  It revealed
some linear trends of $r_{S}-r$ vs.  $r-g$, $g_{S}-g$ vs
$r-g$, $i_{S}-i$ vs  $g-i$, which led to $\approx 0.04$~mag differences
between red and blue stars.  Because of this fact, for determination of
asteroid colours we selected only the solar type SDSS stars (having $0.4 <
g-r < 0.5$~mag and $0.1 < r-i < 0.2$~mag), for which the discrepancies
between both SALT and SDSS photometric systems could be neglected.  The
asteroid magnitudes in each filter were measured with respect to
3-4 SDSS solar colour stars, and the results were averaged.  This way the
uncertainties of the SDSS catalogue were minimised.

On 22 Sep, during 47 minutes, the brightness of the
asteroid, as monitored in our r filter exposures, did not show any
systematic changes. This allowed us to average the measurements, obtained in
r, g, and i filters, and use them to compute colour indices.
The obtained results are: $g-r= 0.59\pm 0.03$~mag, 
$r-i= 0.21 \pm 0.03 $~mag.

\begin{table*}
\caption{Aspect data and observing log}
\label{tab:AspectData}
\begin{center}
\begin{tabular}{lccccrrrrrrr}
\hline \hline
\multicolumn{1}{c}{Date     } &
\multicolumn{1}{c}{Obs. time} &
\multicolumn{1}{c}{r}         &
\multicolumn{1}{c}{$\Delta$}  &
\multicolumn{1}{c}{$\alpha$}  &
\multicolumn{1}{c}{$\lambda$} &
\multicolumn{1}{c}{$\beta$}   &
\multicolumn{1}{c}{$V$}       &
\multicolumn{1}{c}{Mov}       &
\multicolumn{1}{c}{Exp}       &
\multicolumn{1}{c}{Filter}    &
\multicolumn{1}{c}{Telescope} \\
\multicolumn{1}{c}{YYYY-MM-DD} &
\multicolumn{1}{c}{(UTC)}     &
\multicolumn{1}{c}{[AU]}      &
\multicolumn{1}{c}{[AU]}      &
\multicolumn{1}{c}{$[\degr]$} &
\multicolumn{1}{c}{$[\degr]$} &
\multicolumn{1}{c}{$[\degr]$} &
\multicolumn{1}{c}{[mag]}     &
\multicolumn{1}{c}{[''/min]}  &
\multicolumn{1}{c}{[s]}       &
                              &
                             \\
\hline
2008-10-08 & 04:50 -- 05:47 &  1.072     &   0.081     &  26.2    &  42.1     &  -8.8    & 16.7     &  4.3     & 
\begin{minipage}[r]{15mm}1 $\times$ 600 + 3 $\times$ 900 \end{minipage} & Spec     & WHT  \\
2011-09-15 & 19:16 -- 19:35 &  1.1674 & 0.1799 & 24.0 & 327.2 & 12.5 & 18.7 &  1.5  & 60  & r        & SALT \\
2011-09-22 & 21:23 -- 22:13 &  1.1432 & 0.1680 & 31.4 & 324.2 &  9.6 & 18.7 &  1.5  & 60  & g, r, i  & SALT \\
2011-09-26 & 03:56 -- 04:08 &  1.1325 & 0.1636 & 35.0 & 323.1 &  8.0 & 18.7 &  1.5  & 60  & R        & Kuiper\\
2011-10-19 &03:40 -- 04:17 & 1.0654 & 0.1441 & 57.9 &320.7 & -5.7 & 18.9 &  1.6  &60 & V, R & Kuiper\\
2011-10-21 &03:35 -- 03:46 & 1.0605 & 0.1429 & 59.5 &321.0 & -7.1 & 19.0 &  1.7  &60  & R & Kuiper\\
2012-01-19 &06:34 -- 06:44 & 1.0871 & 0.1647 & 47.7 &  74.1 &-36.2 & 19.0 &  3.0  &60 & R & Kuiper\\
2012-01-26 &07:36 -- 07:43 & 1.1078 & 0.1841 & 44.3 & 82.1 &-31.4 & 19.2 &  2.8  &30 & R & Kuiper\\
2012-01-27 &06:49 -- 06:56 & 1.1107 & 0.1871 & 44.0 & 83.1 &-30.7 & 19.2 &  2.8  &60 & R & Kuiper\\
2012-02-23 &05:16 -- 05:27 & 1.2016 & 0.3032 & 40.1 &104.2 &-15.4 & 20.3 &  1.9  &60 & R & Kuiper\\
2012-02-24 &05:00 -- 05:12 &1.2051 & 0.3087 & 40.2 &104.8 &-15.0 & 20.4 &  1.9  &60 & R & Kuiper\\
2012-03-28 & 04:31 -- 04:45 & 1.3233 & 0.5348 & 42.7 &124.1 & -5.2 & 21.8 &  1.5  & 60 & R & Kuiper\\
\hline \hline
\end{tabular}
\end{center}
Note: $r$ and $\Delta$ are the distances of the asteroid from the Sun and the Earth,
respectively, $\alpha$ is the solar phase angle, while $\lambda$ and $\beta$
are the geocentric, ecliptic (J2000) longitude and latitude.
In the next column an average brightness $V$ of the asteroid, as predicted by the
Horizons ephemeris service, is given. Starting from the ninth column, the table
gives the asteroid movement on the sky (Mov), the exposure time (Exp),
and the filters used (the abbrevation \textit{Spec} denotes spectroscopic observations)
\end{table*}

\subsection{Kuiper telescope}
The University of Arizona Kuiper 1.54-m telescope located near Mount Bigelow
in southeastern Arizona was used to obtain V- and R-band photometry of 
2000 $\mbox{FJ}_{10}$ on 9 separate dates between September 2011 and March 2012 (Table~\ref{tab:AspectData}). 
On all 9 nights the asteroid was observed in the R-band. On the night of 2011 October 19 the asteroid was also observed in
the V-band in order to measure its V-R color. The instrument used was the Mont4K, a Fairchild CCD486 $4096\times4097$ CCD,
with a FoV of $9\farcm5 \times 9\farcm5$ and plate scale
of $0.28\arcsec$/pixel when binned $2\times 2$.

All data were reduced with the IRAF software package. The images were
bias-subtracted and flat-fielded with twilight and night sky flat images
using the CCDRED package. The APPHOT package was used to perform
aperture photometry of the asteroid and photometric standard stars. In
order to compensate for variable seeing, the average $FWHM$ was measured for
each image and the photometric aperture was set to a radius of 2$\times
FWHM$. Sky background was measured with a circular ring aperture of radius
20 pixels and width of 10 pixels. The sky aperture was centered on the
position of the measured source. The telescope was tracked at the rate of motion of
the asteroid and the images were shifted and co-added on the motion of the
asteroid.

Photometric V- and R-band reference stars from \citet{Landolt.1992} were observed
at multiple airmasses on each night in order to determine the photometric
zero point and extinction coefficient.

The V- and R-band images were taken in the following sequence: $5\times$R, 
$5\times$V, $5\times$R, $5\times$V, $5\times$R and $5\times$V. 
A V-R color index of~$0.48\pm 0.05$~mag was derived from
the data. The cadence allowed for variability due to the rotation of the
asteroid to be corrected. During the 37 minutes of observations, the
asteroid steadily increased in brightness by $\sim 0.3$ magnitudes.

\subsection{WHT} 
Observations were conducted using the
Intermediate-dispersion Spectrograph and Imaging System (ISIS) mounted on
the 4.2m William Herschel Telescope, La Palma (Table~\ref{tab:AspectData}).  Light from the optical
system was split using the 5300 dichroic (blue cut off/red cut on at 5300
Angstroms) and directed along the red arm of the instrument.  The red arm
uses a red-sensitive 4k$\times$2k pixel RED+ CCD with anti-reflection
coating.  The R158R grating with a slit width of 1 arcsecond were used
producing a dispersion of 1.8 Angstroms per pixel.  The total usable
wavelength coverage from the red arm was 5300 to 10000 Angstroms, however
the S/N degrades rapidly beyond 9000 Angstroms.

To ensure the NEA remained within the slit, non-sidereal tracking was used
with the telescope tracking at the apparent rate of motion of the NEO.  The
position of the NEO was monitored throughout each exposure with manual
corrections to the pointing position applied when necessary.  In total
1 $\times$ 600s + 3 $\times$ 900s exposures were obtained.

\begin{figure}[t!]
\resizebox{\hsize}{!}{\includegraphics[clip]{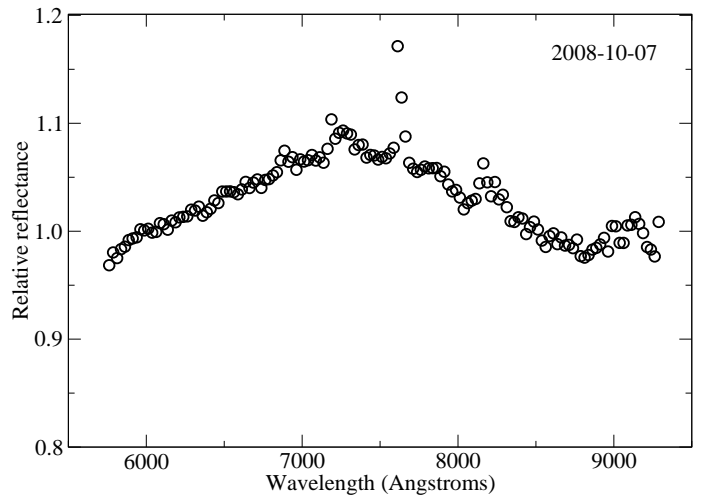}}
\caption{Reflectance spectrum of (190491) 2000 $\mbox{FJ}_{10}$ obtained with the WHT. It is similar
to the spectra of S type asteroids in the SMASS database.}
\label{fig:spectrum}
\end{figure}

Image reduction was performed in the usual manner.  Bias images were
combined and subtracted from the science images.  Flat fielding was
conducted after removal of the spectral profile of the tungsten lamp used to
obtain the flat field images.  The spectra were extracted using the IRAF
task \emph{apall}.  Optimal extraction \citep{Horne1986} was used to improve
the S/N of the extracted spectrum.  Wavelength calibration was achieved
using CuAr-CuNe arc lamp spectra observed at the same pointing position to
account for the effects of flexure.  The solar analog 16 Cyg B was observed
to enable the removal of the solar spectrum from the NEA spectrum. 
Atmospheric correction was performed using tabulated extinction functions
for La Palma \citep{King1985}.  These curves were scaled to the airmass of
the observations of the NEA and solar analog and the corrections applied to
each spectrum.  Asteroid (1) Ceres was also observed and the spectrum extracted in this
manner was consistent with previously published spectra
\citep[e.g][]{BusBinzel2002}. Spectra were normalized and the solar spectrum removed. 
The result for 2000 $\mbox{FJ}_{10}$ is shown in Fig.~\ref{fig:spectrum}.

Chi-squared fitting to the Bus-DeMeo taxonomy
\citep{DeMeo.et.al2009} was conducted.  The spectra were resampled to
produce reflectances at the wavelengths used to define the various taxonomic
types.  The best fit was to an Sq-type, but we note that the 1 micron
absorption band has not been fully sampled, making a formal definition
difficult.

\section{Physical characteristics}
\label{sec:Phys}

\subsection{Rotation period}
\begin{figure}[!t] 
\resizebox{\hsize}{!}{\includegraphics[clip]{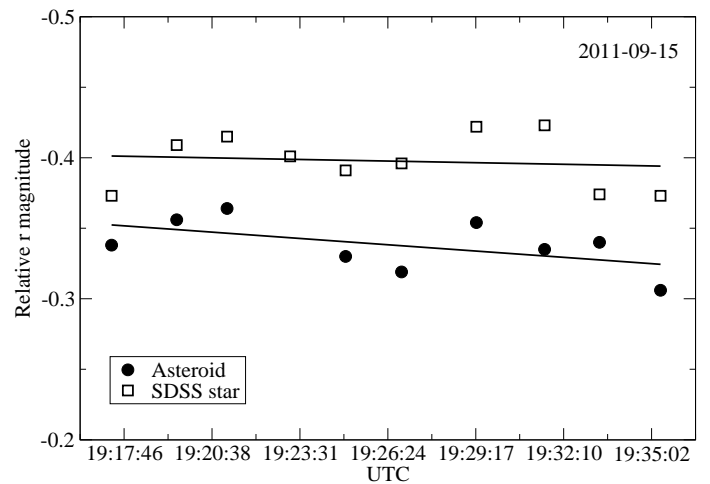}}
\caption{Lightcurve of (190491) 2000 $\mbox{FJ}_{10}$ obtained in the Sloan r filter.  There are no apparent periodic
variations during 20 minutes with a peak-to-peak amplitude greater than
0.05~mag.  The slope in the linear fit is too small to be
significant.  For comparison, brightness variations of one of the SDSS stars
are also presented.  The observed scatter of points is caused by the 
residual effects of the flat fielding procedure.}
\label{fig:lightcurve} 
\end{figure}

As already explained, the specific construction of SALT limits its
typical continuous observing run to 1 hour.  Because of this, it is difficult
to measure brightness variations with periods of several hours or longer. 
In the case of 2000~$\mbox{FJ}_{10}$ there was some possibility that -- due to its small
size -- it is a fast rotator, with a rotation period $\ll$ 2 hr. 
Fig.~\ref{fig:lightcurve} presents one of the lightcurves
obtained during our observations.  It shows relative brightness variations
of the asteroid in the Sloan r filter, and of one of the comparison SDSS stars. 
There are no traces of periodicity in these data with a peak-to-peak
amplitude greater than 0.05~mag.  The observed scatter is probably caused by
imperfect flat fielding rather than statistical noise.  
Assuming that a typical asteroid displays a bimodal brightness variation we can conclude the rotation
period of 2000~$\mbox{FJ}_{10}$ $P$ is longer than twice the time span covered by our
data ($P>36$~min).  
The data obtained on another night with the r filter were analysed in the same way. As it also showed no discernible light variation, we can raise
the lower limit for $P$ to $94$~min.  Of course, there is a small
probability that at the time of our observations the asteroid was visible
close to the pole-on view and its brightness variations were difficult to detect
-- despite its short period.

Similar observations performed on 19~Oct with the Kuiper telescope revealed a
systematic increase of the asteroid's brightness by $\sim 0.3$~mag during 37
min.  This does not contradict our SALT observations as from 22~Sep to
19~Oct the solar phase angle almost doubled from $31\degr$ 
to $58\degr$, while the observer-centred ecliptic latitude changed by 15$\degr$. This could have led to an increase of the lightcurve amplitude.  It is also possible that our SALT observations were performed during maximum brightness, when the
lightcurve for some time could be almost flat.

The fact that on 19~Oct during 37~min the asteroid lightcurve did not show
any extremum can be used to raise further the lower limit for its rotation period
from 94 min to about 2 hours (under the assumption of a typical bimodal
lightcurve). We thus conclude that 2000~$\mbox{FJ}_{10}$ is not a fast rotator.

\subsection{Taxonomy}
\begin{figure}[!t]
\resizebox{\hsize}{!}{\includegraphics[clip]{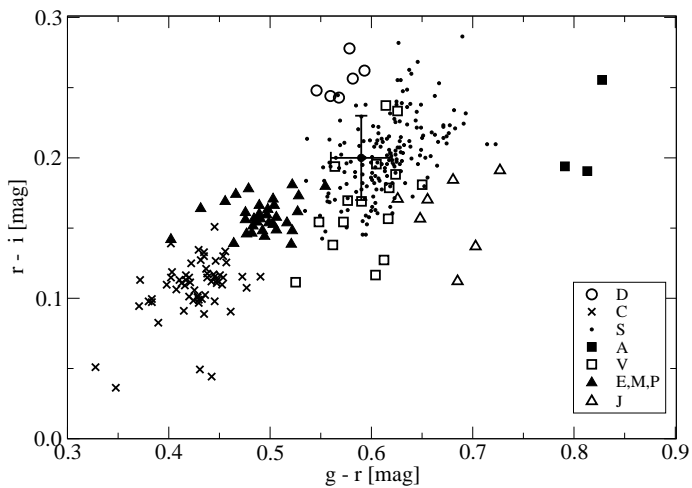}}
\caption{Colour-colour plot for the 316 asteroids whose 
spectra were obtained by the SMASS Survey, based on Fig.~10
from \citet{Ivezic.et.al.2001}. Different taxonomic classes are presented by
different symbols. The colours of 2000 $\mbox{FJ}_{10}$  (filled circle with error bars) are most compatible with an S
type classification, though V, D or E, M, P types cannot be ruled out.}
\label{fig:colour-colour}
\end{figure}

From the photometric standpoint, an asteroid's taxonomy is usually
determined from its colour indices in the Johnson UBVRI system.  Instead of
transforming our Sloan g-r and r-i colours to their B-V and V-R counterparts
we used an analogous classification based on the SDSS magnitudes. 
\citet{Ivezic.et.al.2001} used 316 spectra obtained in the SMASS survey and
convolved them with the SDSS response functions.  As a result they were able
to identify different taxonomic classes on the g-i vs r-i domain
(Fig.~\ref{fig:colour-colour}).  The position of 2000~$\mbox{FJ}_{10}$ on this graph shows
it is most probably an S type, but V, D, or E, M, P types cannot 
be ruled out. This conclusion is supported by the 
$V-R=0.48\pm 0.05$~mag measured with the Kuiper telescope, which is
consistent with an S-type classification \citep{Tholen.Barucci.1989} and in agreement
with the WHT spectrum.

\subsection{Diameter}
The observations of 2000 $\mbox{FJ}_{10}$ over a wide range of phase angles allowed us to
plot its phase curve, and to derive the absolute magnitude as well as the
effective diameter. The R magnitudes measured with the Kuiper telescope
have been transformed to V magnitudes using its V-R colour index.  The r
magnitudes measured with SALT have been transformed to V magnitudes with the
equations given by \citet{Jester.et.al.2005} (for that we
also used the asteroid $g-r$ colour index).

\begin{figure}[!t]
\resizebox{\hsize}{!}{\includegraphics[clip]{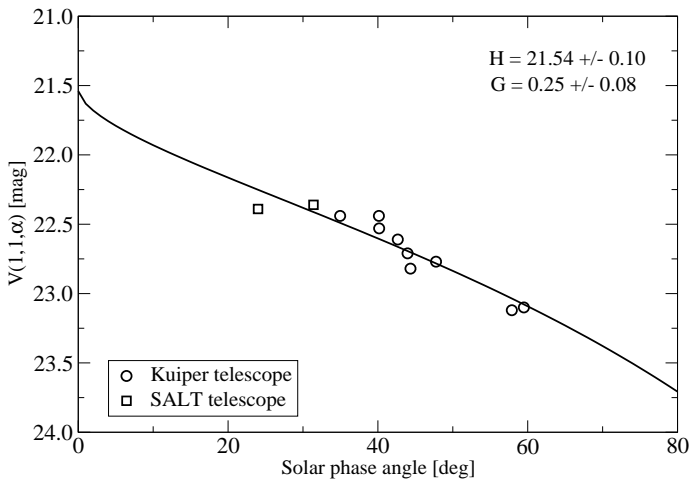}}
\caption{Least-square fit of the H-G relation to the asteroid
phase curve. The scatter of the points is
partially caused by the changing aspect and the lightcurve amplitude.}
\label{fig:phase_curve}
\end{figure}

Figure~\ref{fig:phase_curve} presents a plot of $V(1,1,\alpha)$ magnitudes
versus solar phase angle. The apparent magnitude of the asteroid at each observation
has been corrected to unit distance from the Sun and the Earth. A least square fit with the standard H-G relation 
\citep{Bowell.et.al1989} gives $H=21.54 \pm 0.10$~mag and $G=0.25\pm 0.08$. The quoted uncertainties
were estimated by assuming error bars in $V(1,1,\alpha)$ of $\pm
0.05$~mag. It is worth noting the large discrepancy between $H=20.9$
given by the MPC (based entirely on inaccurate magnitude estimates),
and our result. It confirms a general rule that the MPC absolute
magnitudes derived for NEAs are often underestimated by up to 0.5~mag.

The obtained value of $G=0.25$ is consistent with the average value $G=0.23$
obtained for S type asteroids \citep{Lagerkvist.Magnusson.1990},
further supporting our taxonomic classification of 2000 $\mbox{FJ}_{10}$ as an S type object.

Recently, improved albedo estimates for NEAs in different taxonomic classes were
derived \citep{Thomas.et.al.2011}.  The average geometric albedo for the S
complex was found to be $p_{V}=0.26^{+0.04}_{-0.03}$.  Using this value with
our $H$ magnitude, and the classic formula provided by
\citet{Fowler.Chilemi.1992}, we obtain for 2000~$\mbox{FJ}_{10}$ an effective diameter of
$D_{\mathit{eff}}=0.13\pm 0.02$~km.

\section{Dynamical Modelling}
\label{sec:Orbit}
\begin{table}
\caption{(190491) 2000 $\mbox{FJ}_{10}$ clone classification statistics at different stages during the forward and backward integrations.}
\label{tab:outcomes}
\begin{center}
\begin{tabular}{lcccc}
\hline
\multicolumn{5}{c}{FORWARD}\\ 
\hline
 & \multicolumn{4}{c}{time from integration start (yr)}\\ 
Outcomes    & $+10^{4}$   &     $+10^{5}$    &   $+5 \times 10^{5}$     &    $+10^{6}$ \\ 
\hline
q$>$1.3 AU  &       0       &       0          &         17        &           26     \\  
Amors      &       998      &      283      &        390      &        376   \\        
Apollos      &      1         &      705       &         486      &      425  \\                   
Atens         &        0         &       11       &        106       &     172  \\  
\hline
\multicolumn{5}{c}{BACKWARD} \\ 
\hline
   &  \multicolumn{4}{c}{time from integration start (yr)}  \\
Outcomes  &  $-10^{4}$   &  $-10^{5}$    &   $-5 \times 10^{5}$   &    $-10^{6}$ \\   
\hline                  
       q$>$1.3 AU      &        0       &       1          &         5         &       18  \\
           Amors      &       996    &       979      &         667      &    569  \\
           Apollos       &     3       &         19       &         296      &     333  \\           
           Atens         &        0       &         0        &          31      &       79  \\ 
\hline
\end{tabular}
\end{center}
\end{table}

\begin{figure*}[htb]
\begin{center}
\includegraphics[angle=0,width=6cm]{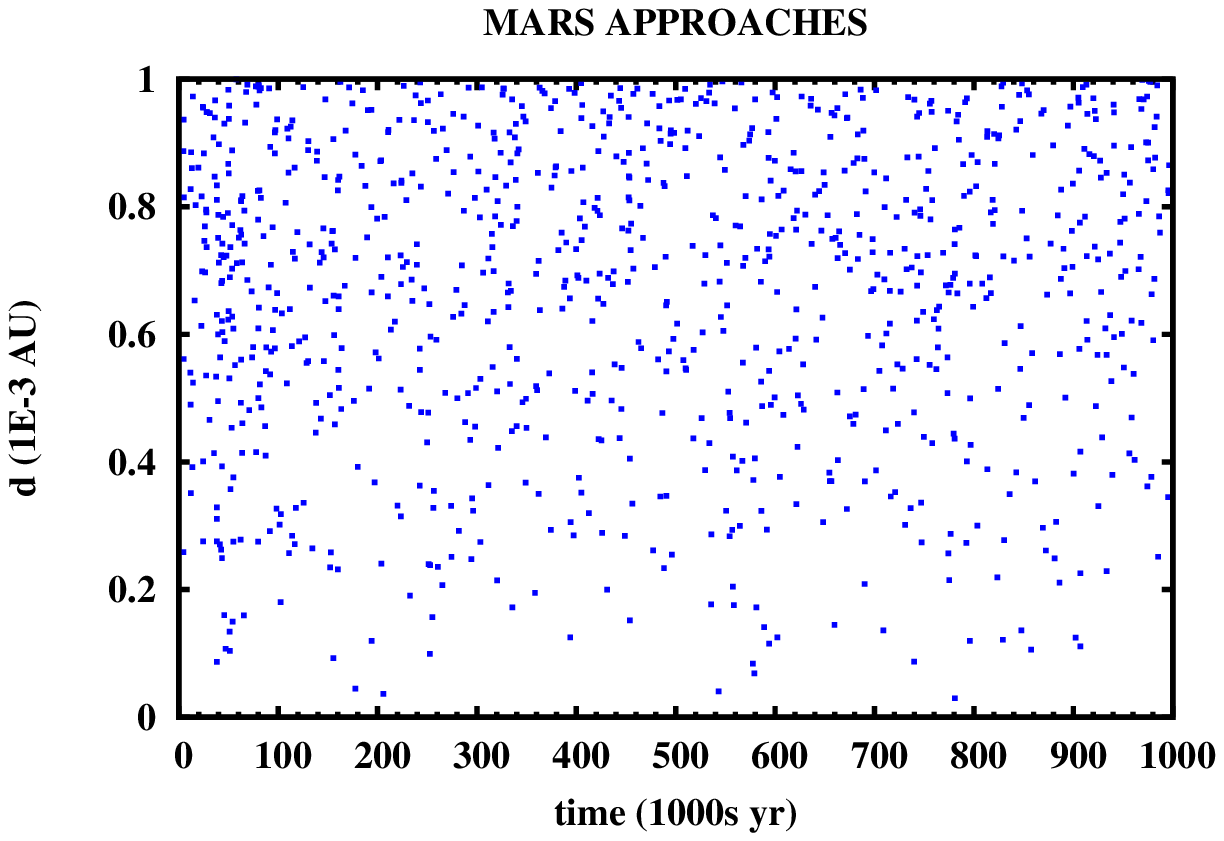}\includegraphics[angle=0,width=6cm]{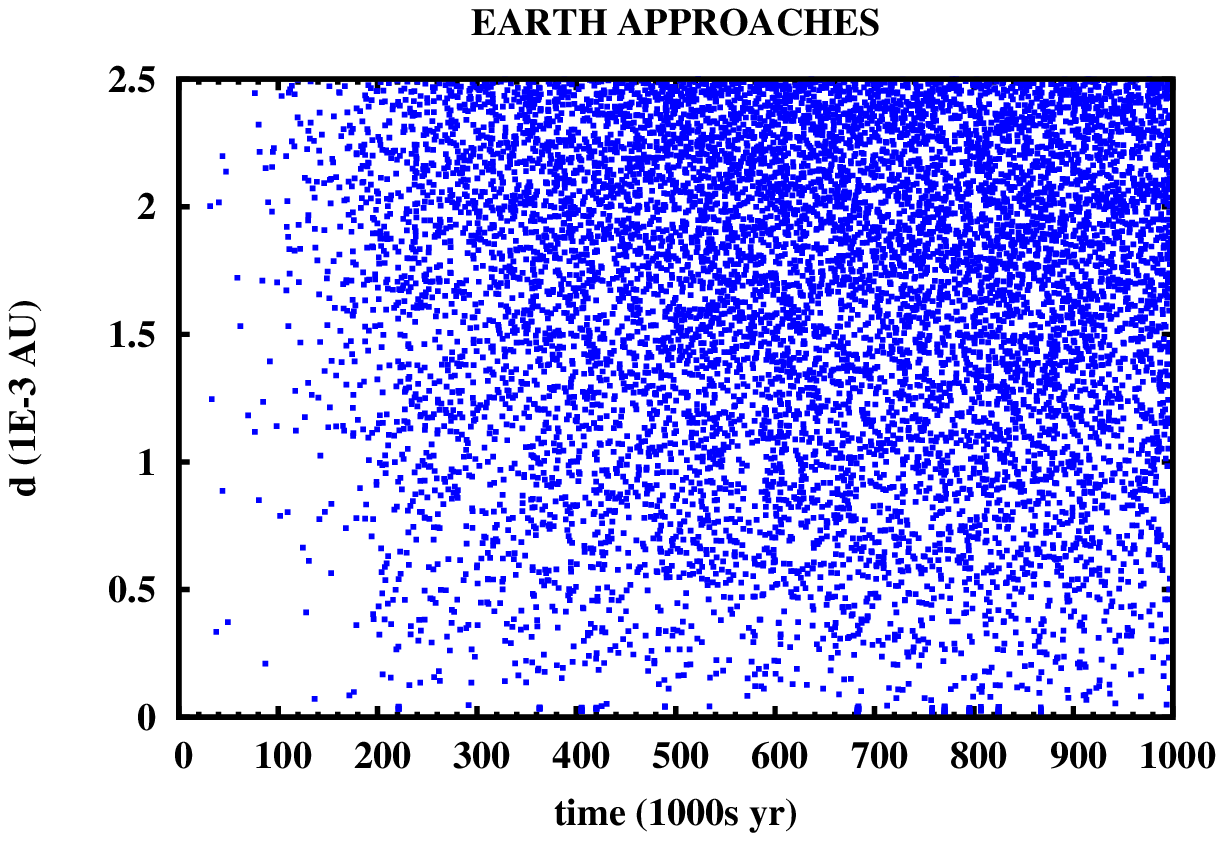}\includegraphics[angle=0,width=6cm]{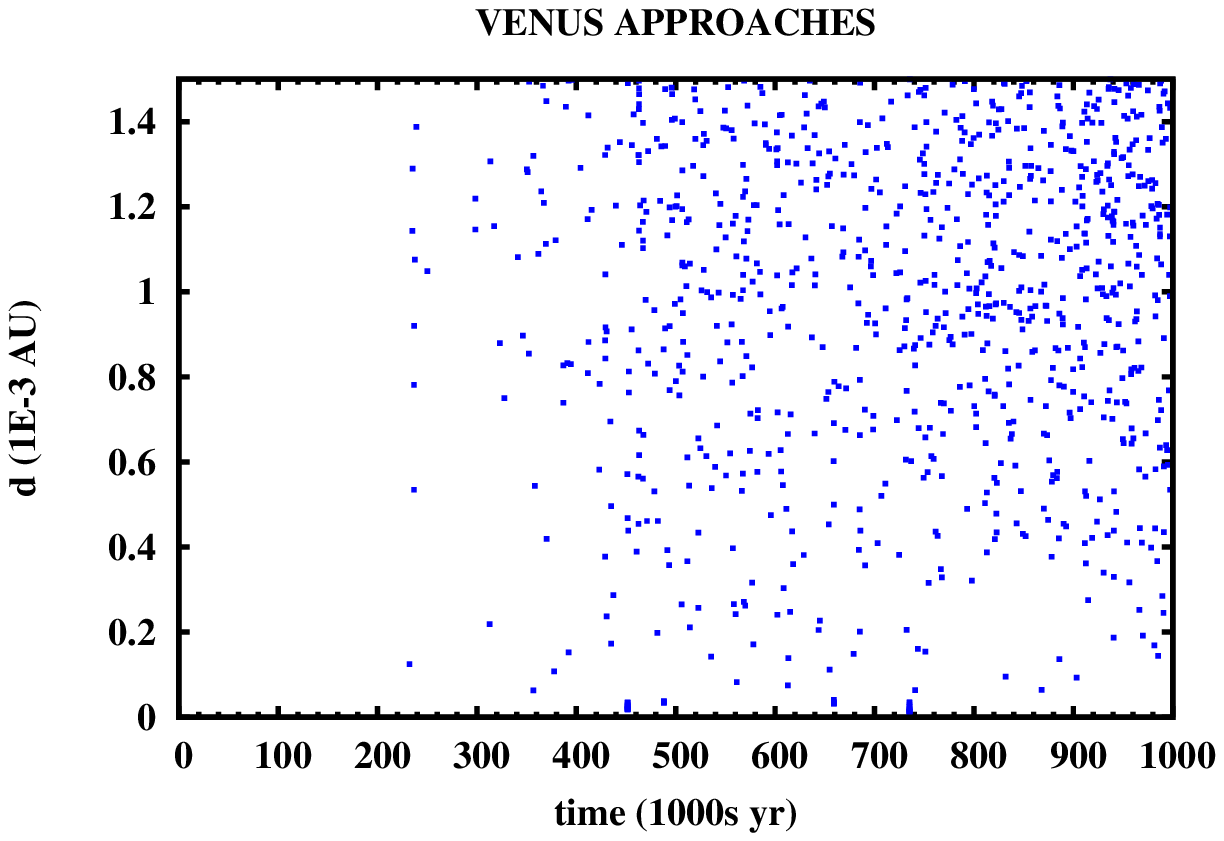}
\includegraphics[angle=0,width=6cm]{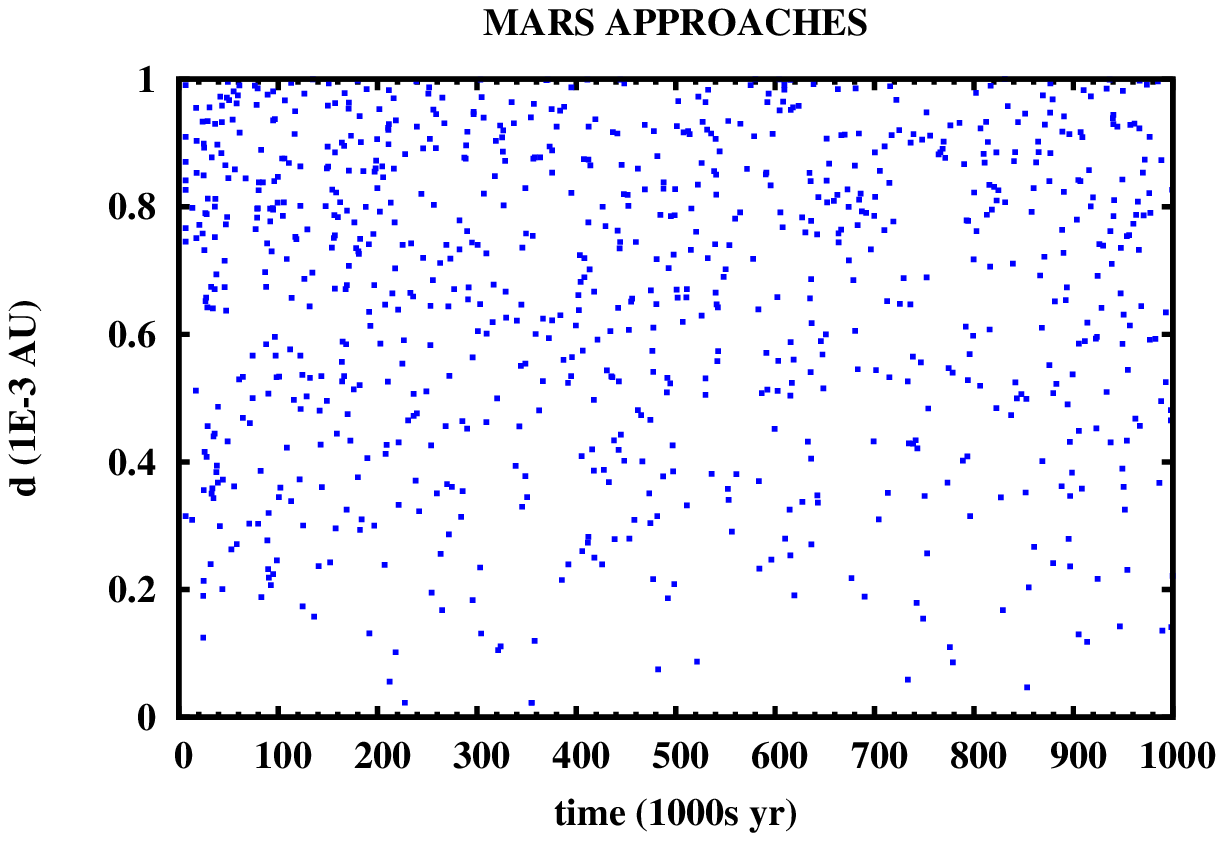}\includegraphics[angle=0,width=6cm]{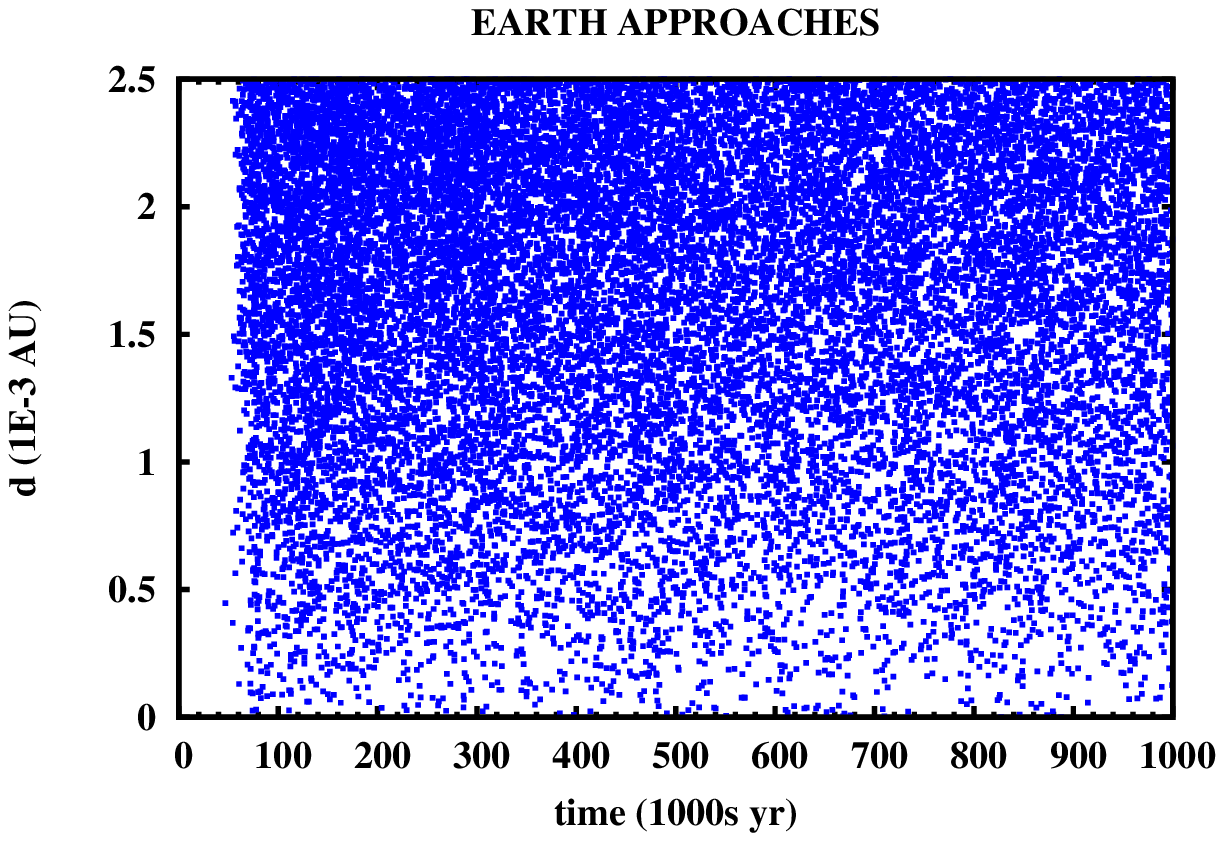}\includegraphics[angle=0,width=6cm]{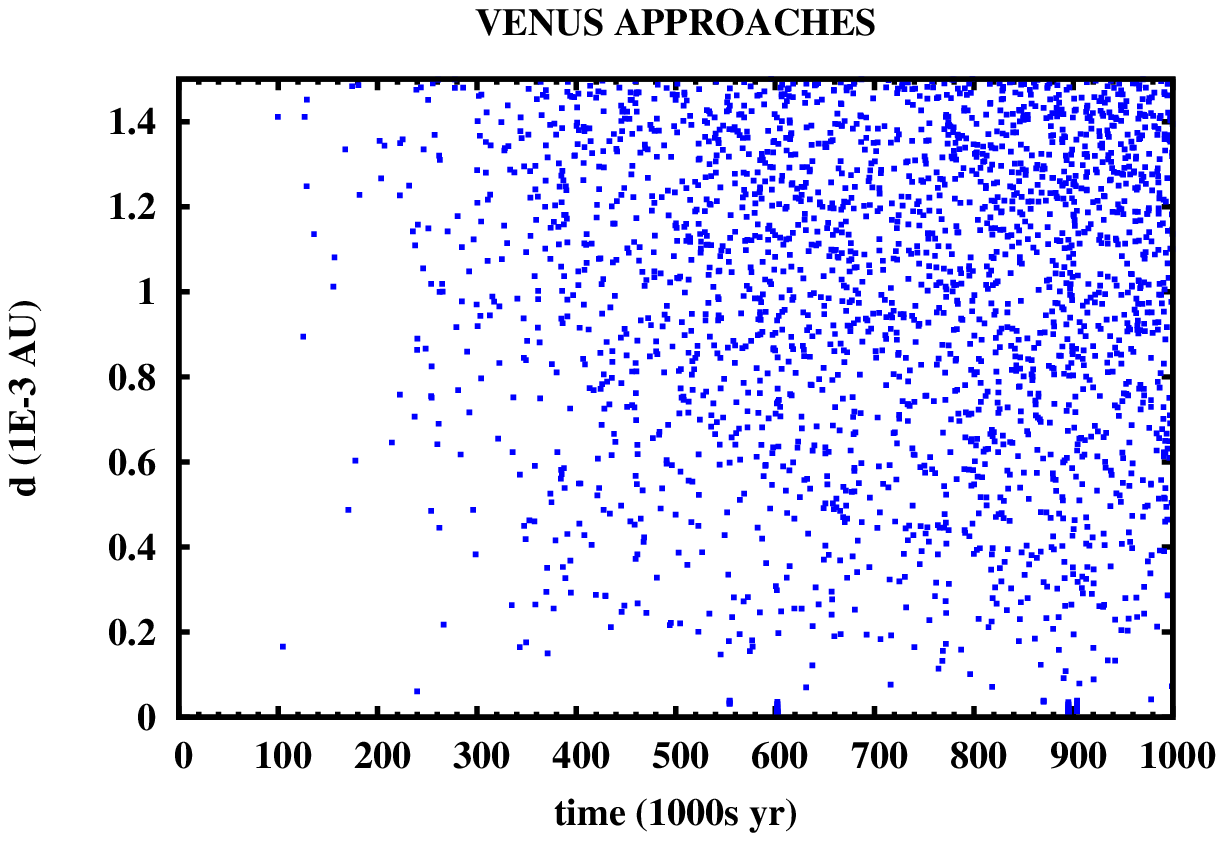}
 \caption[ ]{Minimum distances during encounters with Mars (left), Earth (middle) and Venus (right) as recorded by MERCURY during the backward (top) and forward (bottom) integrations of 1000 clones of (190491) 2000 $\mbox{FJ}_{10}$. \label{fig:closappr}}
 \end{center}
\end{figure*}

\begin{figure*}[htb]
\begin{minipage}[b]{1.0\textwidth}
\centering
\includegraphics[angle=-270,scale=0.7]{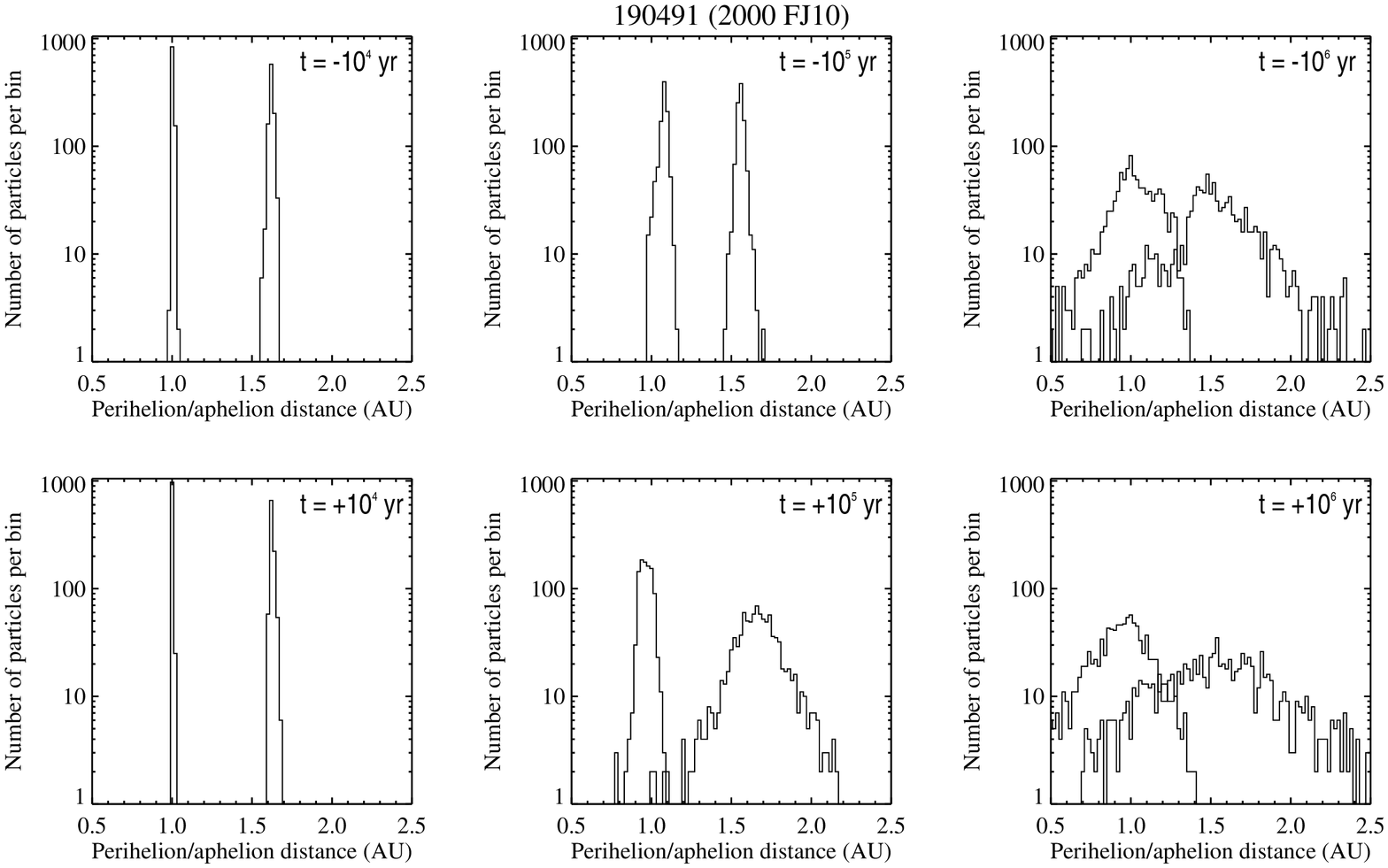}
\end{minipage}
 \caption[ ]{Distribution of clone perihelion and aphelion distances of 1000 clones of (190491) 2000 $\mbox{FJ}_{10}$ at three different times during the backward (top) 
 and forward (bottom) integrations. \label{fig:qr_adist}}
\end{figure*}

\subsection{Method}
To characterise the asteroid's recent orbital evolution and constrain its origin, we have generated 1000 dynamical clones of 190491 by applying the formal state covariance of that 
asteroid for Julian Date 2455600.5 downloaded from {\it AstDys}\footnote{http://hamilton.dm.unipi.it/astdys/} to a six-dimensional 
gaussian random vector \cite[see][ for details]{Duddy.et.al2012}. The clones were then integrated
$10^{6}$ yr in the past and in the future under an 8-planet model of the solar system.
The integrations were carried out using the ÔÔhybridÕÕ scheme which is part of the MERCURY package \citep{Chambers1999}. 
This scheme is based on a second-order mixed variable symplectic (MVS) algorithm; it switches to a Bulirsch-Stoer scheme 
within a certain distance from a massive object. For all the integrations reported here, this distance was 2 Hill radii.
The integration time step was chosen to be 4 days (1/20th of the orbital period of Mercury).
During the integration, MERCURY detected and recorded close approaches with the terrestrial planets.
We caution that, {\bf despite} the time reversibility of the equations of motion in the absence of dissipation, 
backwards integrations will not, in general, provide information of the dynamical history of the asteroid; 
instead, the forward evolution starting from the possible source regions must be considered \citep{Bottke.et.al2002}. 
As we show in the following, however, the case of (190491) 2000 $\mbox{FJ}_{10}$ can be considered exceptional in this sense.
In addition, the Yarkovsky effect \citep{Bottke.et.al2006} was not taken into account; we expect that its contribution to
the dynamical evolution of the asteroid will be negligible compared to those of planetary close encounters and secular perturbations.

\begin{figure*}[htb]
\begin{minipage}[b]{1.0\textwidth}
\centering
\includegraphics[angle=-270,scale=0.7]{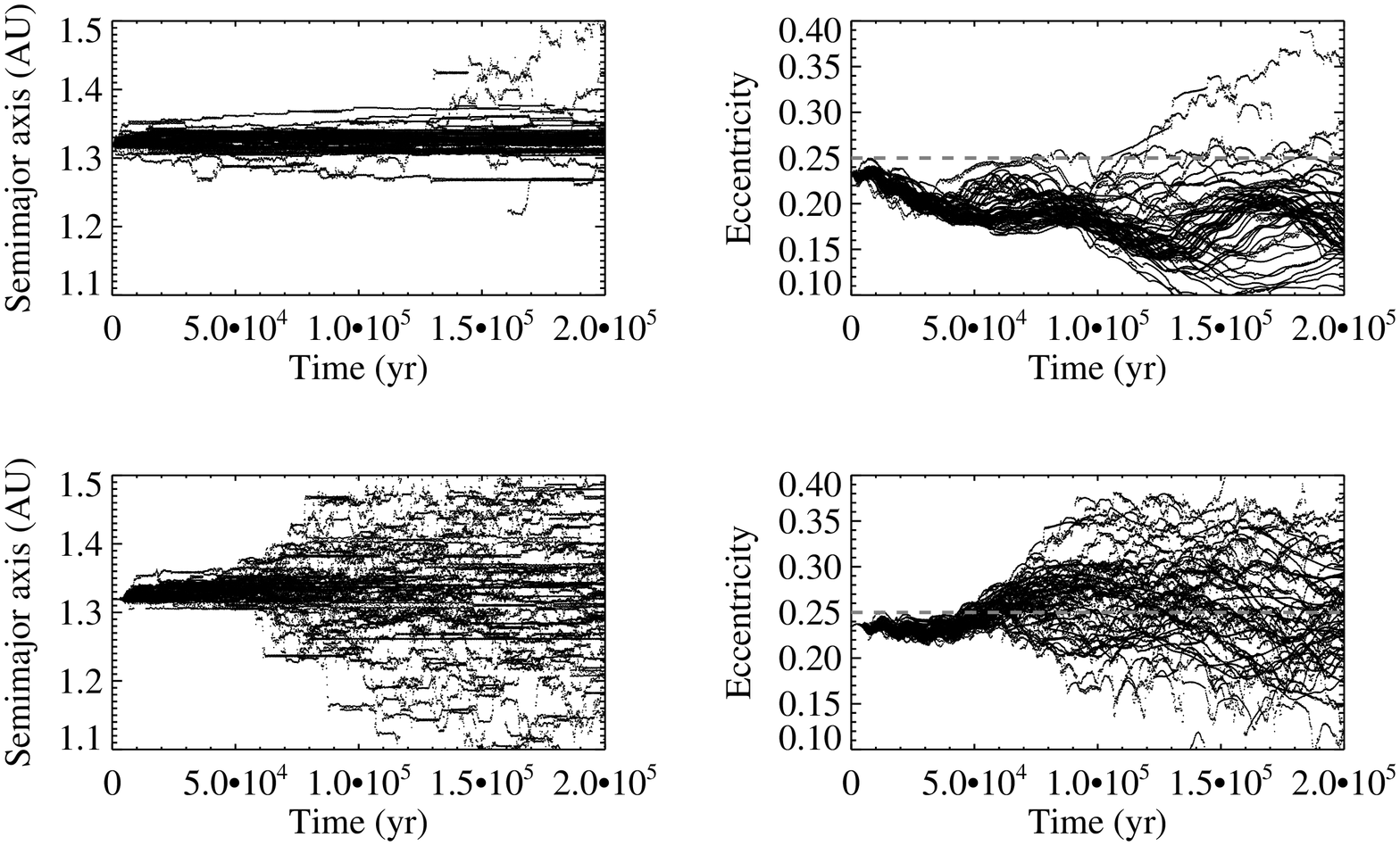}
\end{minipage}
 \caption[ ]{Orbital evolution of asteroid clones $2 \times 10^{5}$ yr backwards (top) and forwards (bottom) from the present. The left panels show the semimajor axis while the right panels show the eccentricity. Only one in every twenty clones is plotted to maintain clarity. The grey dashed line denotes the minimum eccentricity required for an orbit with a semimajor axis of 1.33 AU to cross that of the Earth.\label{fig:a_e_2e5}}
\end{figure*}

\subsection{Dynamical Evolution}
Figures \ref{fig:closappr}, \ref{fig:qr_adist} and Table~\ref{tab:outcomes} show the distribution of the 
close approach distances of the clones to Mars, Earth and Venus as a function of time, the distributions of their 
perihelia and aphelia and their orbital classification statistics. 
Note that we have used the Minor Planet Center definitions for the Amor, Apollo and Aten NEA classifications. 
These are slightly different than those used by other authors \cite[eg][]{Bottke.et.al2002}.

At present the clones of this Amor asteroid are experiencing close approaches with Mars but not with either Earth or Venus 
(left panels of Fig.~\ref{fig:closappr}). Despite the clones' nominal perihelion distance being $\sim$1.01 AU, this situation 
persists $\pm$10 kyr into the integrations. At that time, essentially all clones have $q>1$ AU 
(left panels of Fig.~\ref{fig:qr_adist}) and remain classified as Amors (Table~\ref{tab:outcomes}). 
One likely contributing factor is a 3:2 near-resonance between the orbital periods of 190491 and the Earth, 
which renders close encounters with the planet infrequent if not impossible (see
Section~\ref{sec:Access}).   

 In the backward integrations, close approaches of clones to the Earth occur at a gradually increasing rate (upper middle panel of 
 Fig.~\ref{fig:closappr}) during a period of $\sim$300 kyr. Their perihelia remain at or above 1 AU at $-100$ kyr (upper middle panel of 
 Fig.~\ref{fig:qr_adist}); consequently they retain their Amor classification (Table~\ref{tab:outcomes}). By that time, only 15 clones 
 have experienced close encounters with the Earth. In the forward integrations, the onset of Earth encounters is abrupt
 and occurs between +50 and +100 kyr. At +100 kyr most of the clones have perihelia $<$ 1 AU and have become Apollos (lower 
  middle panel of Fig.~\ref{fig:qr_adist}; Table~\ref{tab:outcomes}). This difference between the outcomes of the forward and backward integrations is likely due to the secular evolution of the eccentricity. Fig.~\ref{fig:a_e_2e5} shows the time series
  of $a$ and $e$ $\pm 2\times 10^{5}$ yr from the present for one in every 20 clones. Although the evolution of individual clones is inherently chaotic, we note a statistical trend 
  towards smaller values of $e$ in the past and larger values of $e$ in the future. The timescale of variation (50+ kyr) leads us to suspect that it
  is related to one or more of the secular eigenmodes of the solar system \citep{Laskar1990}. The ``critical'' value of $e$ required to reduce
  $q$ below 1 AU is $0.25$ for a mean $a$ value of 1.33 AU. Once $e$ increases past that critical value - indicated by the gray dashed line - in the forward integrations,
  Earth encounters become possible (see Fig.~\ref{fig:closappr}) and, as a result, the scatter in $a$ and $e$ (bottom panels) increases significantly. In the backward integrations, the critical value for $e$ is never reached for all but a few cases.  
  
  This result likely reflects a predictable event of the 
  asteroid's dynamical evolution. In other words the real asteroid will begin encountering the Earth sometime in the interval 
  50-100 ky in the future and, as a consequence, become an Apollo asteroid.

The final orbital distribution of the clones is shown in the right panels of Fig.~\ref{fig:qr_adist} and the last column of 
Table~\ref{tab:outcomes}. Assuming Poissonian statistics (ie $\sigma=\sqrt{N}$), the asteroid is more likely to be an Amor than 
not an Amor (3-$\sigma$ level) at $t=-1$ Myr. In the forward integrations, we can make the weaker statement that the most 
likely out of the four possible outcomes we examined is an Apollo classification. However, this result is only significant at the 
2-$\sigma$ level when compared to that of an Amor classification. The final perihelia and aphelia of the clones appear to be more 
widely dispersed for the forward than for the backward integrations. Finally, we note that the forward integrations produced a
significant fraction of Atens and that two of the clones collided with the Sun, a common end result of NEA dynamical evolution 
\citep{Farinella.et.al1994}.
\subsection{Probability of collision with the Earth}
The distribution of close approaches as a function of distance can be used to estimate the collision probability with the Earth.
To do this, we first converted the recorded distances $q_{i}$ and speeds ${\rm v}_{i}$ at closest approach to impact parameters $b_{i}$ and velocities at infinity
${\rm v}_{\infty,i}$ through the expressions \citep{Opik1976}:
\begin{equation}
{\rm v}_{\infty,i} = \sqrt{{\rm v}^{2}_{i} - 2 \mu/q_{i}}
\label{Eq:vinfty}
\end{equation}
\begin{equation}
b_{i}=q_{i} \sqrt{1 + \frac{2 \mu} {q_{i} {\rm v}^{2}_{\infty,i}}}
\label{Eq:imp_param}
\end{equation}
where $\mu$ denotes the product of the planet's mass with the gravitational constant $G$.

\begin{figure}[htb]
\begin{center}
\includegraphics[angle=0,width=8cm]{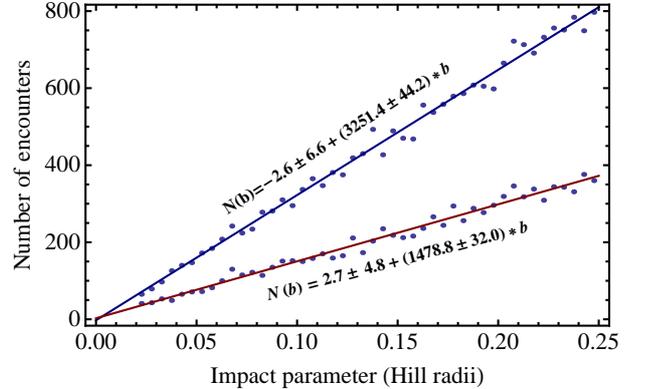}
 \caption[ ]{Linear fit to the number of close encounters to the Earth as a function of the impact parameter $b$ for the forward (upper curve) 
 and the backward (lower curve) integrations. \label{fig:N_vs_b}}
 \end{center}
\end{figure}

Then we collected the $b_{i}$ into bins of width $\Delta b_{i} = 0.005$ in units of the Earth's Hill radius ($\simeq 0.01$ AU) 
and fitted linear laws to the data for the backward and forward integrations separately, expecting that 
\begin{equation}
N(b_{i})=N(b_{i}<b<b_{i}+\Delta b_{i}) \sim b_{i} \Delta b_{i}\mbox{.}
\end{equation}
 Figure~\ref{fig:N_vs_b} shows that the frequency of deep close encounters ($\ll 1 R_{H}$) is higher in the future than in the past,
consistent with the higher dispersion of future over past clones as found earlier in this Section. The expected number of collisions with the Earth 
per Myr is $N_{c}=\int_{0}^{b_{E}} N(b) db$ where $b_{E}$ is calculated from Eq.~\ref{Eq:imp_param} by setting $q=R_{E}$, Earth's radius
and assuming ${\rm v}_{\infty, E} = <{\rm v}_{\infty,i}> \simeq 5.8$ km $\mbox{s}^{-1}$. This evaluates to $\sim0.11$ and $\sim0.08$ for the forward 
and backward integrations respectively. Hence the probability of impact of (190491) 2000 $\mbox{FJ}_{10}$ with the Earth in the interval $\left[-10^{6},10^{6}\right]$ yr is 
$\sim 2 \times 10^{-4}$. 

\section{Synthesis: The Origin of 2000 $\mbox{FJ}_{10}$}
\label{sec:Origin}

\citet{Bottke.et.al2002} found that $\sim85\%$ of all NEOs with $H<22$ originate in either the inner (IMB; $a < 2.5$ AU) or central (CMB; $2.5$ AU $<a < 2.8$ AU) Main Belt. About half ($\sim53\%$) of existing Amors originate in the inner Belt alone. They are delivered 
in the terrestrial planet region via the 3:1 mean motion resonance with Jupiter, the $\nu_{6}$ secular resonance and close encounters with Mars 
\cite[see also][]{Binzel.et.al2004}. The S-complex taxonomic classification deduced from our observations is consistent with this premise.
The dynamical simulations show that it has been an Amor for at least 100 kyr from the present. The gradual loss of determinacy in the eccentricity evolution (Fig.~\ref{fig:a_e_2e5}) allows us to extend the validity of this statement for up to a few 100s of kyr in the past.
Since only a few clones attained perihelia $> 1.3$ au at the end of both the forward and backward integrations, we conclude that, if it arrived from the asteroid belt, it likely did so $>1$ Myr ago. A small but statistically significant difference between the outcomes of the full 1 Myr runs in the past and in the future suggests that the real asteroid is currently evolving from the Amor to the Apollo dynamical class. However, given the caveats in interpreting backwards integrations mentioned in Section 5.1, we believe this
conclusion to be tentative. To better quantify the likelihood of different scenarios, it would be necessary to apply methods such
as that of \citet{Bottke.et.al2002}.

We also note that $>$ 95\% of clones in both forward and backward integrations maintained an inclination of $ < 15^{\degr}$, indicating that it is unlikely to have originated in the high inclination 
Hungaria and Phocaea families. If it originated within either the IMB or the CMB \citep[source regions of 75\% of Amors according to][]{Bottke.et.al2002} it may be
a former member of a family dominated by S-type asteroids, the most populous of which are the Flora and Eunomia families \citep{Zappala.et.al1995}.

\section{Accessibility from the Earth}
\label{sec:Access}

\begin{figure}[htb]
\centering
\includegraphics[angle=0,width=8cm]{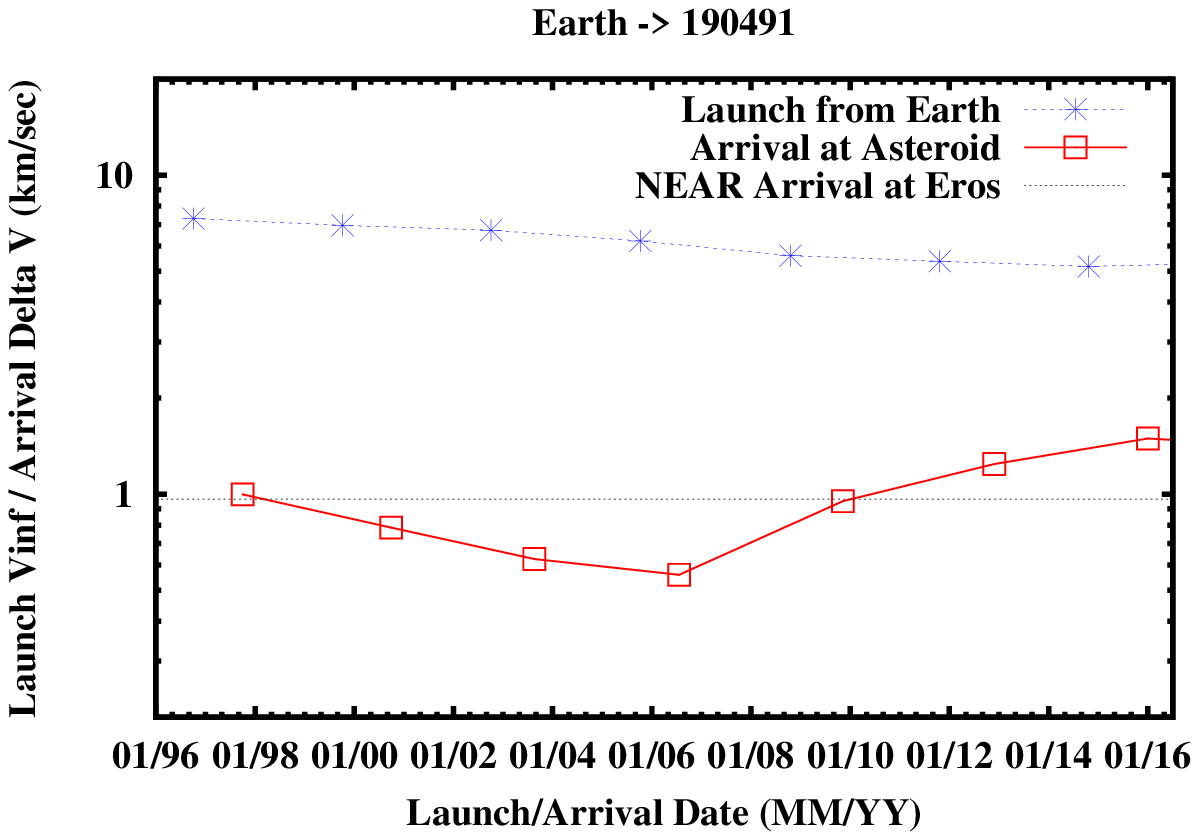}
\includegraphics[angle=0,width=8cm]{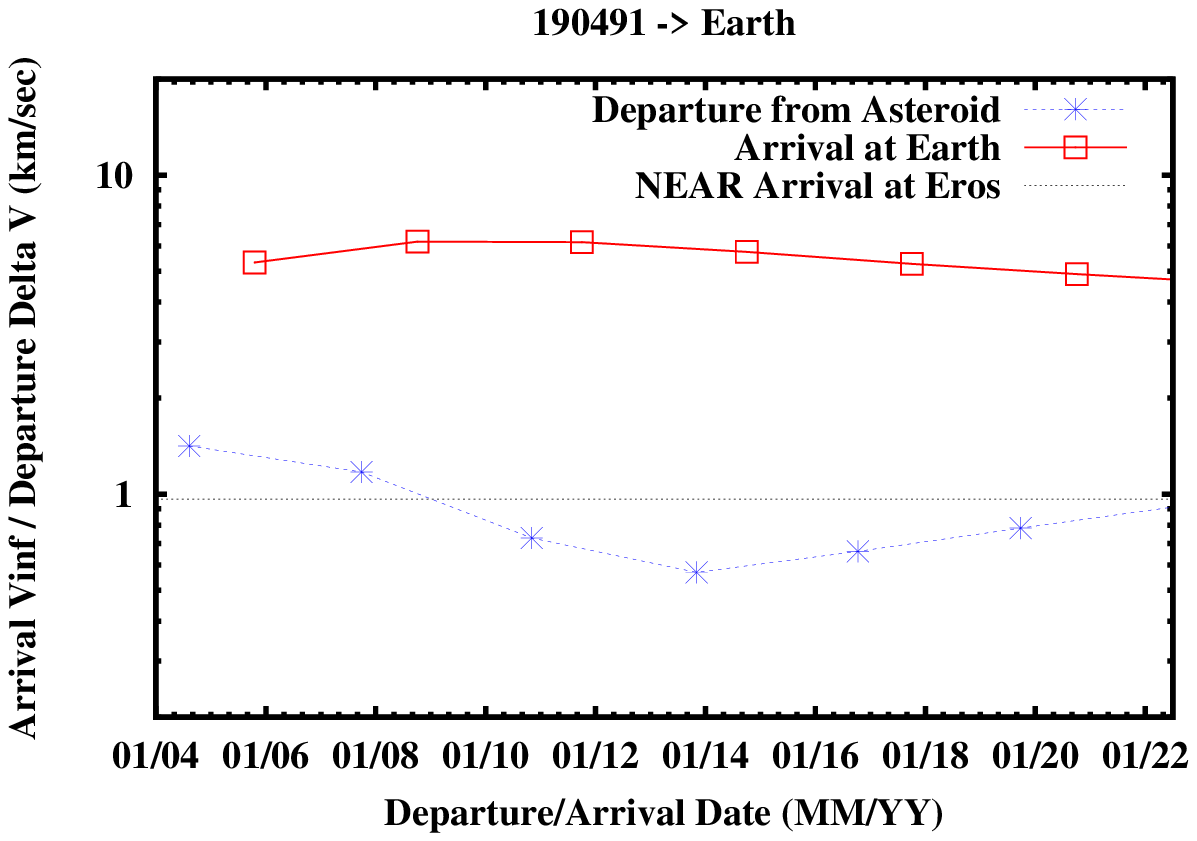}
 \caption[ ]{Top: Departure $v_{\infty}$ and $\Delta V$ required to match speeds with 190491 at arrival. The $\Delta V$ expended 
 by NEAR upon arrival at 433 Eros is indicated. Bottom: Return leg.}
\label{fig:dv}
\end{figure}

To quantify the accessibility of the asteroid from the Earth, we constructed two way
(Earth - Asteroid - Earth) Type II keplerian arcs based on the Gauss method as in
\citet{Bate.et.al1971}. Arrival and departure dates were determined by minimising the $\Delta V$ at the asteroid. 
We considered that a launch/return window existed when this quantity was equal to or less than that expended
by the NEAR spacecraft in 1999 to rendezvous with 433 Eros in early 2000 \citep[$\sim$0.965 km $\mbox{s}^{-1}$;][]{Dunham.et.al2002}.
We found that consecutive launch windows for the asteroid are spaced $\sim$3 years apart (upper panel of Fig.~\ref{fig:dv}) and that
the same holds for return windows (lower panel of Fig.~\ref{fig:dv}). Typical one-way trip times are $\sim$1 year with 
a one-year wait at the asteroid before insertion into the Earth return trajectory.  

\begin{figure*}[htb]
\centering
\includegraphics[angle=0,width=8cm]{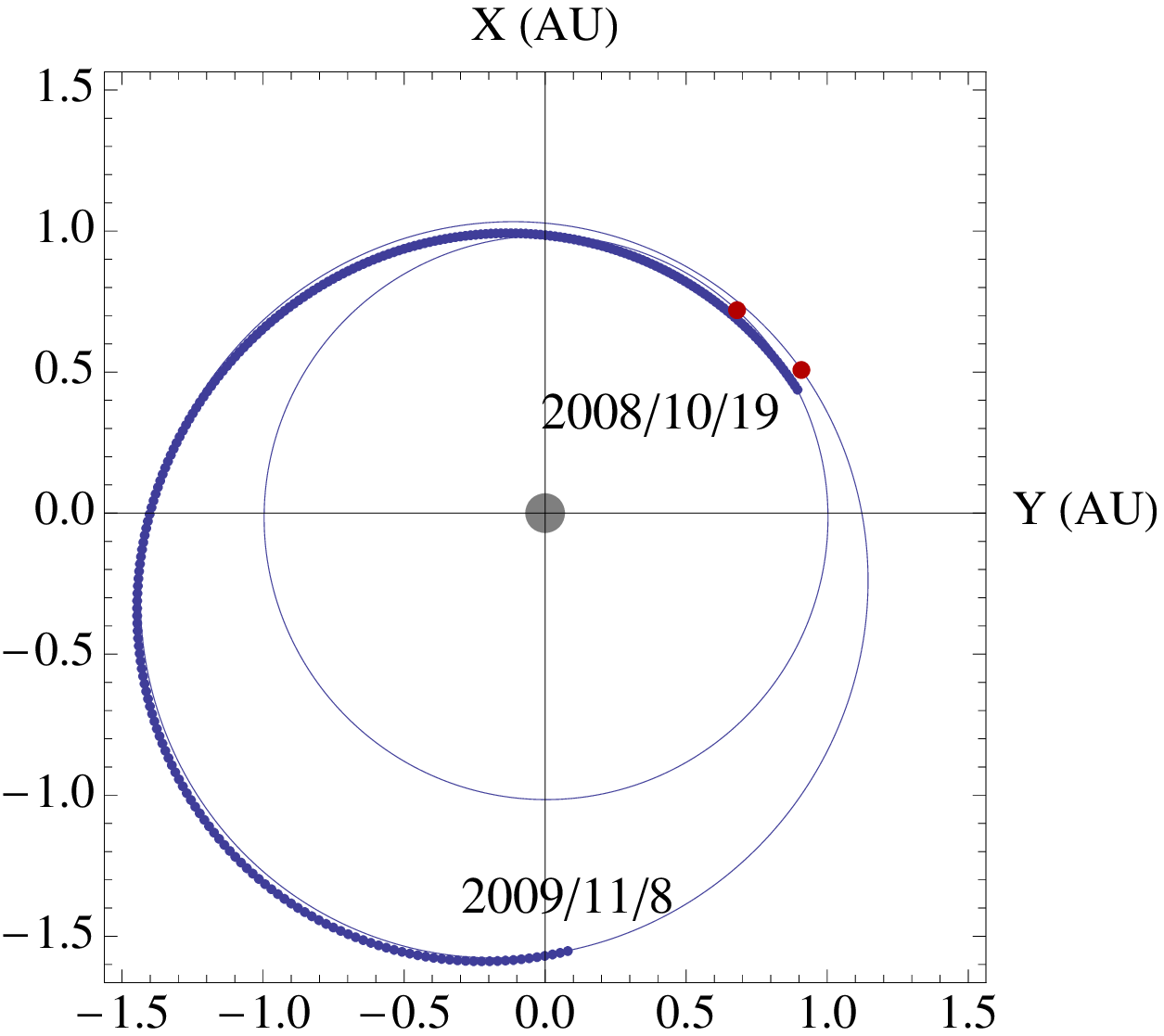}\includegraphics[angle=0,width=8cm]{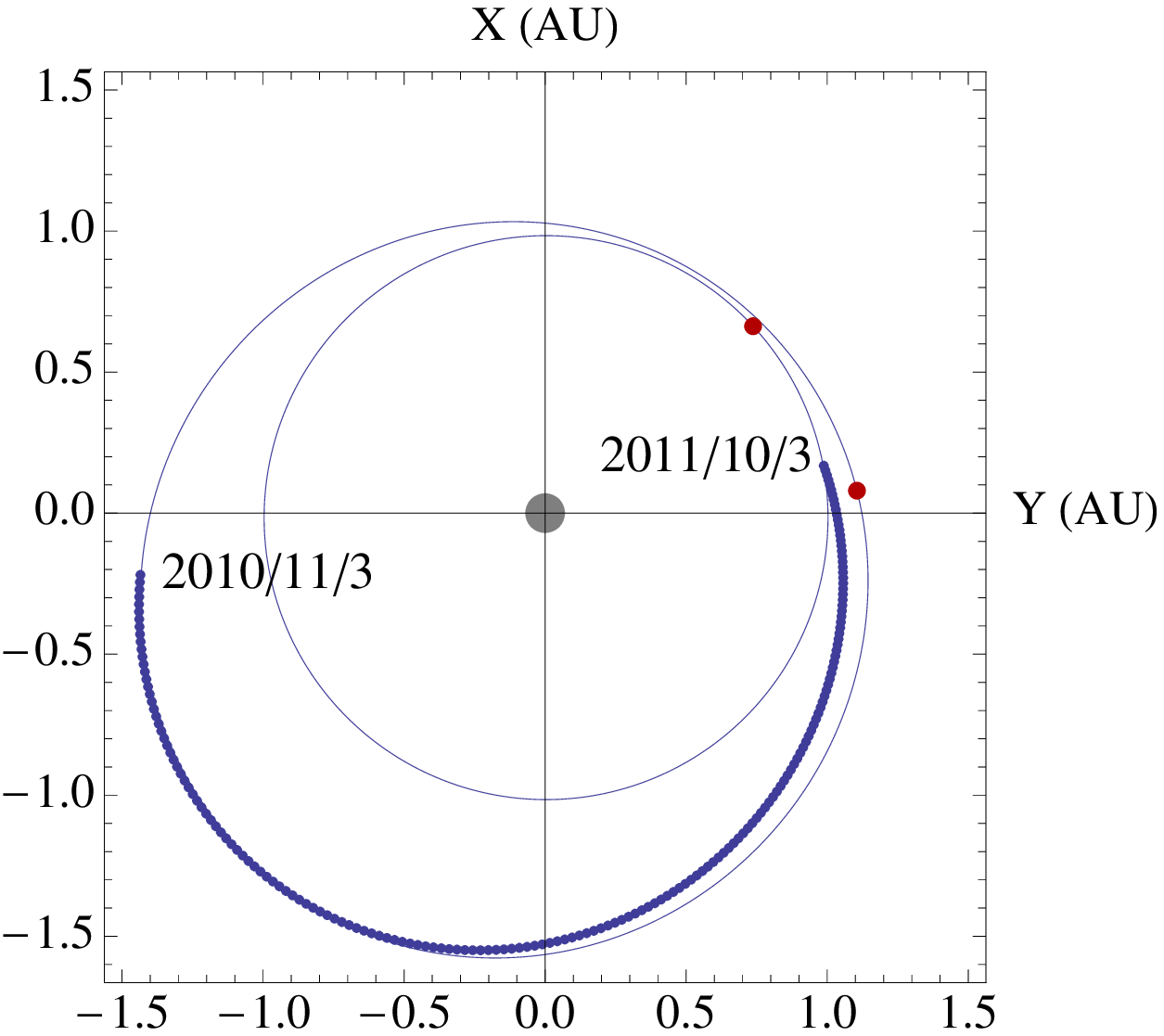}
 \caption[ ]{Round trip to 190491 using Type II two-impulse transfers. Left: outbound leg. The locations
 of the asteroid at departure and the Earth at arrival are denoted by the red disks. Right: Return leg.
 The red disks now indicate the location of the Earth at departure and that of the asteroid at Earth arrival.}
\label{fig:roundtrip}
\end{figure*}

A set of example trajectories generated by our code are shown in 
Fig.~\ref{fig:roundtrip}. Departure from Earth occurs in Autumn 2008 
($v_{\infty}$ = 5.5 km $\mathrm{s}^{-1}$) and arrival the 
following Autumn ($\Delta V$ = 0.95 km $\mathrm{s}^{-1}$). Departure one year later 
($\Delta V$ = 0.75 km $\mathrm{s}^{-1}$) brings the hypothetical spacecraft  back to 
the Earth in October 2011 ($v_{\infty}$ = 6.3 km $\mathrm{s}^{-1}$). It is important to 
note for the discussion that follows that departure from and return to the 
Earth occurs when the asteroid is nearby.

Interestingly, during the period 2000-2100 AD favourable launch windows
occur in two groups, the first in 2000-2012 and the second in 2047-2059. 
Corresponding favourable return windows span the periods 2010-2020 and 2058-2070 respectively. On
those occasions rendezvous with the asteroid requires $<1$ km $\mathrm{s}^{-1}$ 
of $\Delta V$ and a similar amount for departure and return to the Earth.
The total $\Delta V$ for arrival at, and departure from, the asteroid  is $< 2$  km $\mathrm{s}^{-1}$ on
three round-trip opportunities in the period 2052-2061. 

\begin{figure*}[htb]
\centering
\includegraphics[angle=0,width=5cm]{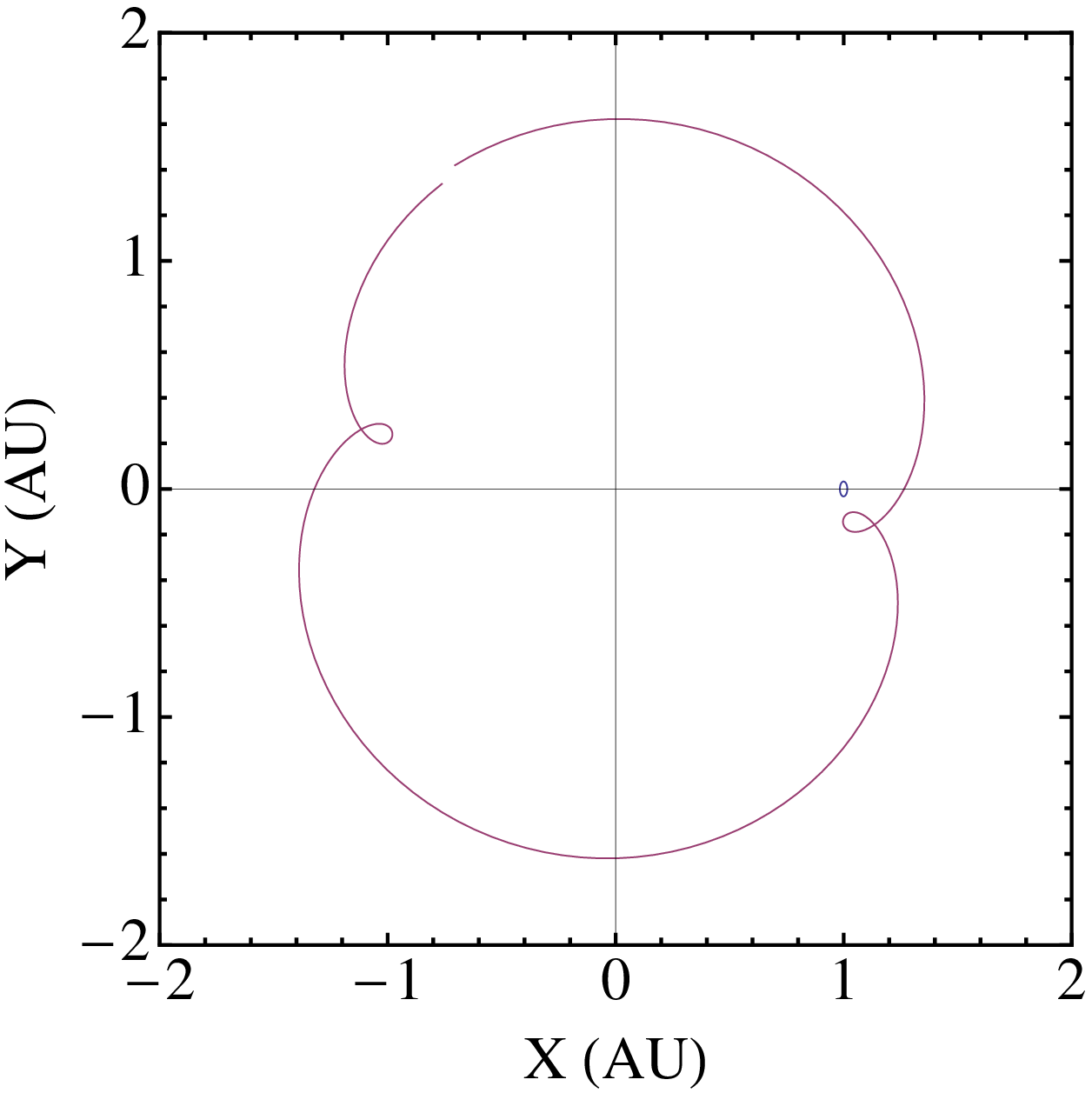}\includegraphics[angle=0,width=5cm]{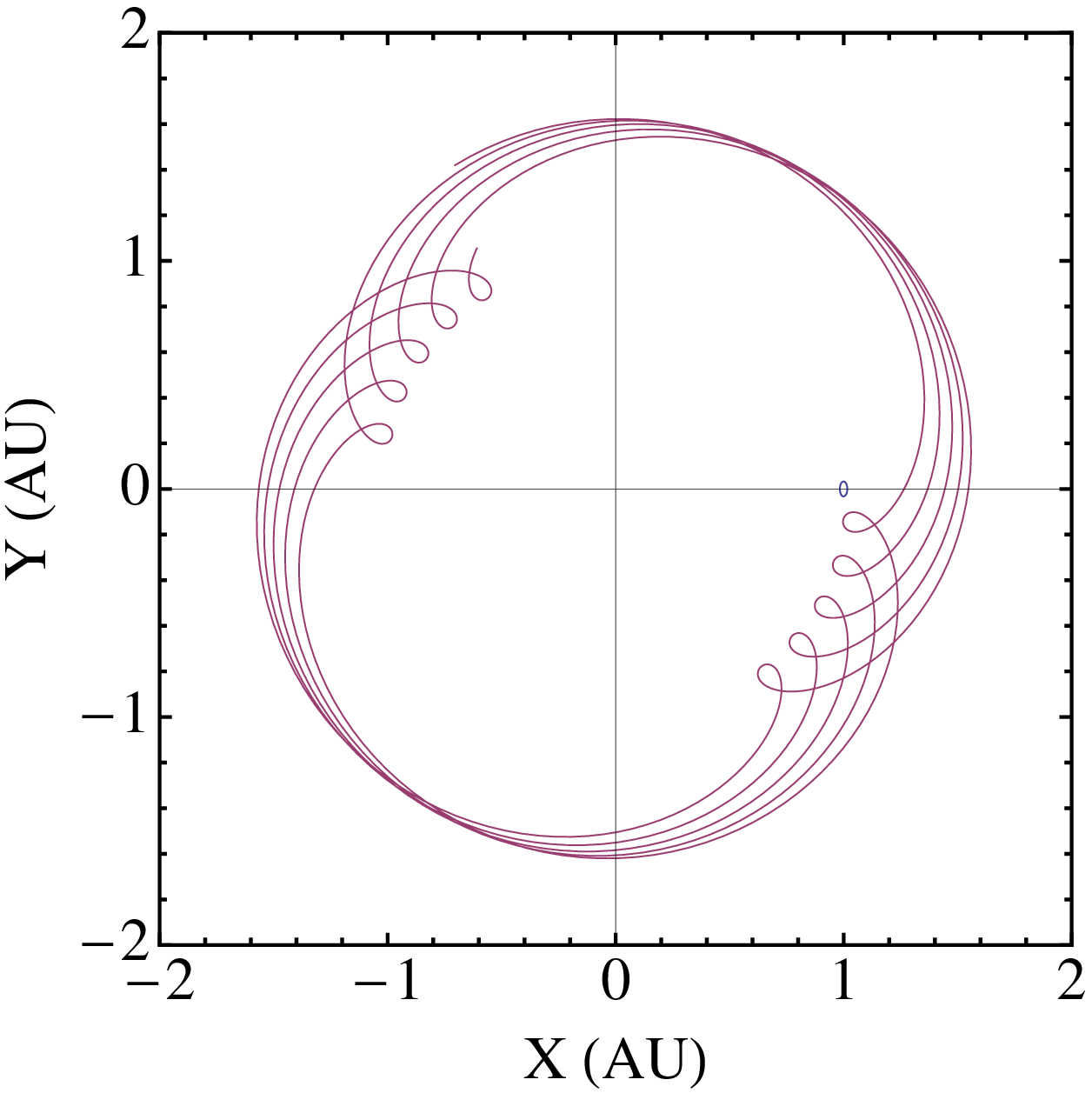}\includegraphics[angle=0,width=5cm]{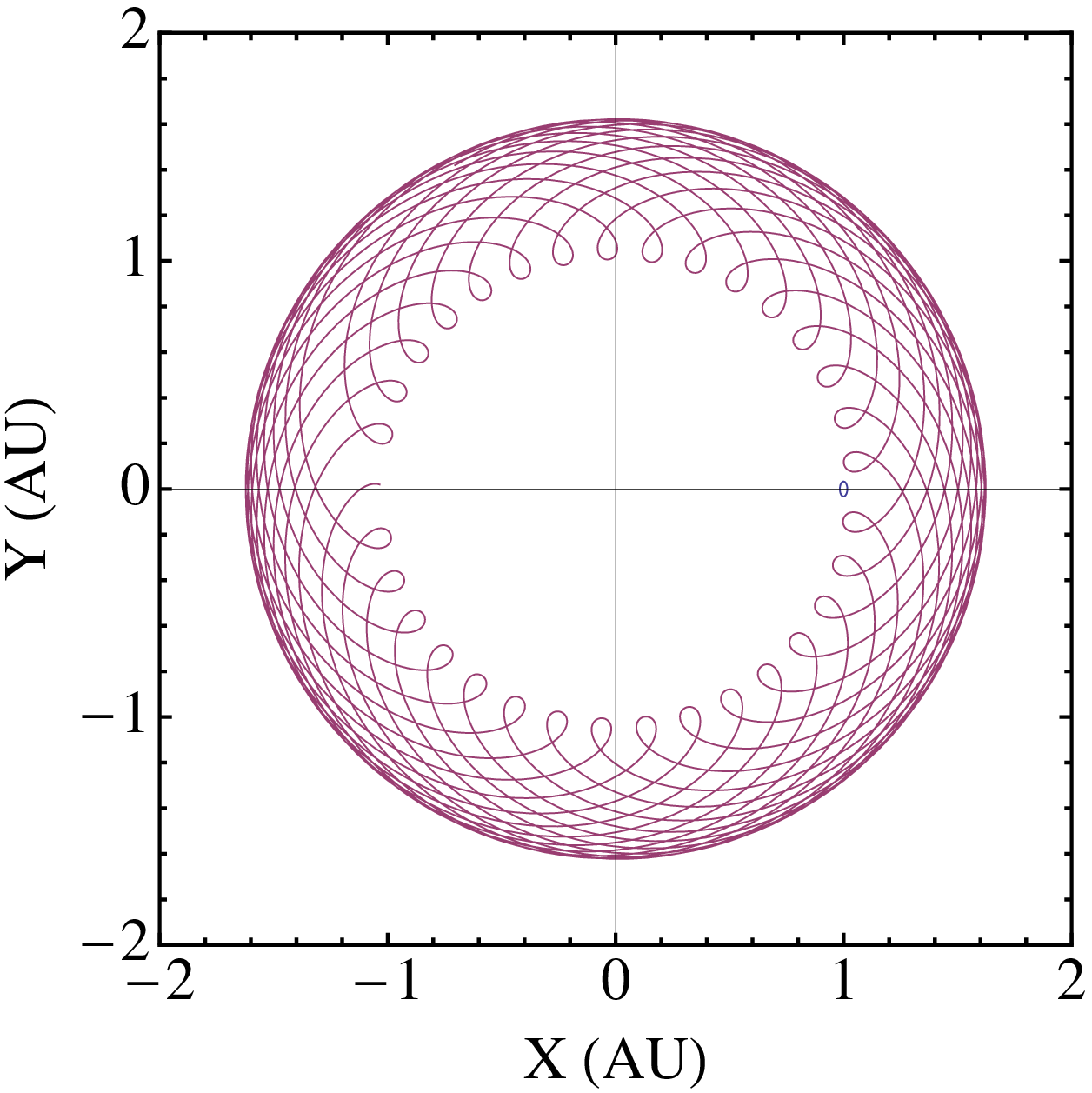}
 \caption[]{The trajectory of the asteroid in a frame rotating with the Earth's mean motion
 over periods of 3yr (left panel), 15 yr (middle panel) and 50 yr (right panel).
 The  small ellipse at (1,0) represents the trajectory of the Earth in this frame, due to its orbital eccentricity ($\sim 0.017$).}
\label{fig:rotframe}
\end{figure*}

We find that this is the result of a 3:2 near-resonance between the orbital period of the asteroid and that of the Earth. This is best demonstrated 
if one views the asteroid's trajectory in a frame that rotates with the Earth's mean angular velocity around the Sun (Fig.~\ref{fig:rotframe}).
Due to the near-resonance the asteroid traces out a pattern with respect to the Earth completing one revolution every 3 Earth years (left panel). 
The two loops correspond to the asteroid passing through the pericentre of its orbit. As of 2011 (left panel), one of those loops lies close to the Earth. 
Because the resonance is not exact, the pattern slowly precesses in a clockwise direction so that, after 15 years (i.e.~2026) close approaches to the Earth are 
no longer possible (middle panel). Half a precession period later (18,000 d or 50 yr) the original configuration is recovered and close approaches 
to the Earth become possible again (right panel). In fact, querying the asteroid's ephemeris using the MPC online tool\footnote{http://www.minorplanetcenter.net/iau/MPEph/MPEph.html} 
shows that the two closest approaches of the asteroid to the Earth ($\Delta \simeq 0.1$ AU) for the remainder of the 21st century occur in 2058 and 2061.

\section{Conclusions}
\label{sec:Summary}

The Amor NEA (190491) 2000~$\mbox{FJ}_{10}$ is one of the most accessible spacecraft
targets.  Its effective diameter of $D_{\mathit{eff}}=0.13\pm 0.02$~km
places it in the transition zone between gravitationally bound rubble-piles
and monolithic bodies, held together by their internal strength \citep{Asphaug.et.al2002}.

2000~$\mbox{FJ}_{10}$ belongs to the S-complex of evolved asteroids, making
it scientifically less interesting than the primitive objects that are usually the targets of robotic sample-return missions.  
On the other hand, its relatively large size, slow rotation, and accessibility from the Earth can make it a source of minerals
and elements important for industry and a suitable target for a piloted mission \citep{Abell.et.al2009}.

Due to its Earth-like orbit, during the next 10 years 2000~$\mbox{FJ}_{10}$ will make 
several close approaches to the Earth which can be used to determine its shape and spin axis orientation. 
Until 2020, there will be favourable apparitions of decreasing quality every 3 years when
the asteroid will be brighter than $V=21$~mag. This is rather
uncommon among NEAs.

\begin{acknowledgements}
Astronomical research at the Armagh Observatory is funded 
by the Northern Ireland Department of Culture, Arts and Leisure (DCAL). 
TK and MB were supported by the Narodowe Centrum Nauki grant
N~N203~403739. The authors wish to acknowledge the SFI/HEA Irish Centre for High-End
Computing (ICHEC) for the provision of computational facilities and support.
Based on observations made with the William Herschel Telescope operated on the island of La Palma by
the Isaac Newton Group in the Spanish Observatorio del Roque de los Muchachos of the Instituto de Astrofisica de Canarias.
Some of the observations reported in this paper were obtained with the Southern African Large Telescope (SALT). A.~G. acknowledges support from
the National Research Foundation (NRF) of South Africa.
\end{acknowledgements}
\bibliographystyle{aa}
\bibliography{aa_2012_20156}
\end{document}